# Hydrogen and the Abundances of Elements in Gradual Solar Energetic-Particle Events

**Donald V. Reames**

Institute for Physical Science and Technology, University of Maryland, College Park, MD 20742-2431 USA, email: dvreames@umd.edu

**Abstract** Despite its dominance, hydrogen has been largely ignored in studies of the abundance patterns of the chemical elements in gradual solar energetic-particle (SEP) events; those neglected abundances show a surprising new pattern of behavior. Abundance enhancements of elements with $2 \leq Z \leq 56$, relative to coronal abundances, show a power-law dependence, *versus* their average mass-to-charge ratio $A/Q$, that varies from event to event and with time during events. The ion charge states $Q$ depend upon the source plasma temperature $T$. For most gradual SEP events, shock waves have accelerated ambient coronal material with $T < 2$ MK with decreasing power-laws in $A/Q$. In this case, the proton abundances agree rather well with the power-law fits extrapolated from elements with $Z \geq 6$ at $A/Q > 2$ down to hydrogen at $A/Q = 1$. Thus the abundances of the elements with $Z \geq 6$ fairly accurately predict the observed abundance of H, at a similar velocity, in most SEP events. However, for those gradual SEP events where ion enhancements follow positive powers of $A/Q$, especially those with $T > 2$ MK where shock waves have reaccelerated residual suprathermal ions from previous impulsive SEP events, proton abundances commonly exceed the extrapolated expectation, usually by a factor of order ten. This is a new and unexpected pattern of behavior that is unique to the abundances of protons and may be related to the need for more streaming protons to produce sufficient waves for scattering and acceleration of more heavy ions at the shock.

Keywords: Solar energetic particles · Solar system abundances · Coronal mass ejections · Solar flares

# 1. Introduction

Hydrogen is, by far, the most abundant element in typical astrophysical settings, including the Sun, where the pattern of element abundances is believed to have changed little during the 4.6 billion years since its formation.  As elements evaporate from the solar photosphere up into the corona, those that are ionized, *i.e.* with first ionization potential (FIP) < 10 eV, are preferentially enhanced by a factor of about four relative to the elements with FIP > 10 eV that are initially neutral atoms.  Once in the 1-MK corona, all elements are highly ionized, and at altitudes where the plasma becomes collisionless, ion acceleration by electromagnetic fields can be sustained.  The effects of those fields on acceleration and transport depends upon the magnetic rigidity of the ions, which, when compared at a constant velocity, varies as the ion mass-to-charge ratio $A/Q$.  Thus it is not surprising that the effects of acceleration and transport on the abundances of solar energetic particles (SEPs) are found to vary as power laws in $A/Q$.  The surprise is that scientists have advanced this picture of abundance variations of the elements for decades without including the abundance of hydrogen.  Do protons ever fit this abundance pattern?

Early measurement of abundances, averaged over the large SEP events we now call "gradual" events, began to show clear evidence of the FIP effect (*e.g.* Webber, 1975; Meyer, 1985; Reames, 1995, 2014) especially for the elements from He or C through Fe.  Once typical element ionization states $Q$ first became available (Luhn *et al.* 1984), event-to-event variations were found to be power-law functions of $A/Q$ for elements with atomic numbers $6 \leq Z \leq 30$ by Breneman and Stone (1985), but H was also omitted from more recent similar studies by Reames (2016a, 2016b, 2018b).  Even the inclusion of He is rather complicated (Reames, 2017c).  Power-law behavior of abundances in "impulsive" SEP events was proposed by Reames, Meyer, and von Rosenvinge (1994) using abundances of elements with $2 \leq Z \leq 26$ and the extension to high $Z$ shows an especially strong behavior as the third power of $A/Q$ for $2 \leq Z \leq 82$ (Reames, 2000, 2017a; Mason *et al.*, 2004; Reames and Ng, 2004; Reames, Cliver, and Kahler, 2014a).

The terms "impulsive" and "gradual" have come to refer to the dominant physical process of particle acceleration in these SEP events as evidence for two mechanisms has evolved (Reames, 1988, 1999, 2013, 2015, 2017a; Gosling, 1993). The evidence of unique physics in the small impulsive SEP events was first shown by huge 1000-fold enhancements of $^3$He/$^4$He (Serlemitsos and Balasubrahmanyan, 1975; Mason, 2007; Reames, 2017a). While $^3$He/$^4$He $\approx 5 \times 10^{-4}$ is found in the solar wind, even the early measurements found SEP events with $^3$He/$^4$He $\approx 1.5 \pm 0.1$. A few events even have $^3$He/H > 1 (Reames, von Rosenvinge, and Lin, 1985). The events were associated with streaming non-relativistic electrons and type III radio bursts (Reames, von Rosenvinge, and Lin, 1985; Reames and Stone, 1986). In the most complete theory, the streaming electrons produce electromagnetic ion-cyclotron (EMIC) waves in resonance with the gyrofrequency of $^3$He to produce the unique enhancements (Temerin and Roth, 1992; Roth and Temerin, 1997). Further complicating the behavior, power-law enhancements of elements, also 1000-fold, between He and the heavy elements up to Pb (Reames, 2000, 2017a; Mason *et al.*, 2004; Reames and Ng, 2004; Reames, Cliver, and Kahler, 2014a) were found and were better explained by particle-in-cell simulations of magnetic reconnection (Drake *et al.*, 2009), which produce the power-law dependence on *A/Q* from rigidity dependence during acceleration. Impulsive SEP events are associated with narrow coronal mass ejections (CMEs) related to solar jets (Kahler, Reames, and Sheeley, 2001; Bučík *et al.*, 2018) driven by magnetic reconnection involving open field lines.

     In gradual or long-duration SEP events, particles are accelerated at shock waves driven out from the Sun by fast, wide CMEs (Kahler *et al.* 1984). These shock waves accelerate SEPs (Lee, 1983, 2005; Zank, Rice, and Wu, 2000; Cliver, Kahler, and Reames, 2004; Ng and Reames, 2008; Gopalswamy *et al.* 2012; Lee, Mewaldt, and Giacalone, 2012; Desai and Giacalone, 2016; Reames, 2017a), requiring shock speeds above $500 - 700$ km s$^{-1}$ (Reames, Kahler, and Ng, 1997), and accelerating ions over an extremely broad region of the heliosphere (Reames, Barbier, and Ng, 1996; Rouillard *et al.*, 2012; Cohen, Mason, Mewaldt, 2017; Reames, 2017b). Generally, the shock waves sample the $1 - 2$ MK coronal plasma (Reames, 2016a, 2016b, 2018b). Power-law dependence in gradual events can come from rigidity-dependent scattering of ions during transport (Parker, 1963; Ng, Reames, and Tylka, 1999, 2001, 2003, 2012; Reames, 2016a, 2016b), which results in dependence upon *A/Q* when ion abundances are compared at the same particle velocity. If Fe scatters less than O in transit, for example, Fe/O

will increase early in an event and be depleted later on; this dependence on time and space traveling along the field $\boldsymbol{B}$ also becomes longitude dependence because of solar rotation.

However, SEP abundances become complicated because shock waves can also reaccelerate residual ions from a seed population that includes pre-accelerated ions from the ubiquitous small impulsive SEP events, complicating the SEP abundance story. Mason, Mazur, and Dwyer (2002) noted the presence of modest enhancements of $^3$He in large SEP events that would otherwise be called gradual and the complexity of the seed population was soon realized (Desai *et al.*, 2003; Tylka *et al.*, 2005; Tylka and Lee, 2006; Bučík *et al.*, 2014, 2015, 2018; Chen *et al.*, 2015; Reames 2016a, 2016b). Thus gradual SEP events sometimes exhibit the abundance enhancement patterns of impulsive SEP events, which we can recognize from power-law enhancements increasing with $A/Q$ and source plasma temperatures of ≈3 MK (Reames, Cliver, and Kahler, 2014a, 2014b, 2015; Reames, 2019a, 2019b) which actually dominate 24% of gradual events (Reames 2016a, 2016b). The $^3$He/$^4$He ratio is also enhanced in these events (Reames, Cliver, and Kahler, 2014b) but its extreme variation with energy and from event to event (Mason 2007) makes Fe/O a more stable and reliable indicator of impulsive material. Hence we prefer Fe/O rather than $^3$He/$^4$He to identify impulsive SEP events and reaccelerated impulsive ions (Reames, Cliver, and Kahler, 2014a). Recently, the abundance patterns of H (Reames, 2019b) and $^4$He (Reames 2019a) have been found for impulsive SEP events.

Protons have been studied in many aspects of gradual SEP events (*e.g.* Reames, 2017a), including acceleration (Lee, 1983, 2005; Zank, Rice, and Wu, 2000; Ng and Reames, 2008), transport (Parker, 1963; Ng, Reames, and Tylka, 1999, 2001, 2003, 2012), time variations (Reames, Ng, and Tylka, 2000; Reames, 2009), and energy spectra (Reames and Ng, 2010), but not for FIP or power-law abundance patterns. If we seek guidance from the case of the solar wind, we find variations are seen in H/He with solar-wind speed and phase in the solar cycle (Kasper *et al.*, 2007). However, while the solar wind is coronal in origin and does display a FIP effect (*e.g.* Schmelz *et al.* 2012), it is clear that SEPs have a different source than the solar wind (Mewaldt *et al.*, 2002; Desai *et al.*, 2003; Reames, 2018a) that may be related to differing open- and closed-field origins (Laming, 2009, 2015) of the two coronal samples (Reames, 2018a, 2018b). For our pur-poses, the SEP reference coronal abundances (derived from averaged gradual SEP events) and the solar photospheric abundances are provided in Appendix A. We define an element "enhancement" to be its observed abundance, normalized to O, divided by its corresponding reference abundance, similarly normalized.

A recent study of impulsive SEP events extended the power-law dependence of abundances on $A/Q$ from elements with Z>6 down to protons at $A/Q$=1 (Reames, 2019b). It was found that observed proton enhancements generally agreed well with the extrapolated value for small events with no evidence of shock acceleration. However, for most of the larger events, especially those associated with fast CMEs, observed proton intensities were a factor of order ten larger than expected. These impulsive SEP events are associated with solar jets where narrow CMEs can be fast enough to drive shock waves. It seems that the proton excess can be a signature of the presence of shock acceleration in impulsive SEP events. In all impulsive SEP events, abundance enhancements increase as positive powers of $A/Q$.

To what extent can we organize the abundance of H within the scheme of the other elements in gradual SEP events? We consider the gradual events studied and listed by Reames (2016a) using observations made by the *Low-Energy Matrix Telescope* (LEMT) on the *Wind* spacecraft, near Earth (von Rosenvinge *et al.*, 1995; Reames *et al.*, 1997; see also Chapt. 7 of Reames, 2017a). LEMT primarily measures elements He through Pb in the region of 2 – 20 MeV amu$^{-1}$. The element resolution of LEMT up through Fe is shown in detail by Reames (2014). LEMT resolves element groups above Fe as shown by Reames (2000, 2017a). Unlike other elements, the protons in LEMT are limited to a small interval sampled near ≈2.5 MeV bounded by the front-detector (dome) threshold (von Rosenvinge *et al.*, 1995; Reames *et al.*, 1997).

Throughout this text, whenever we refer to the element He or its abundance, we mean $^4$He, unless $^3$He is explicitly stated.

## 2. Power-Laws in *A/Q*

The theory of diffusive transport provides support for our expectation that element enhancements in gradual events will vary approximately as power laws in $A/Q$. It is com-mon to expect that the scattering mean free path $\lambda_X$ of species X depends upon as a power law on the particle magnetic rigidity $P$ as $P^\alpha$ and upon distance from the Sun $R$ as $R^\beta$ so that the expression for the solution to the diffusion equation (Equation C3 in Ng, Reames, and Tylka, 2003 based upon Parker, 1963) can be used to write the enhancement of element X relative to oxygen as a function of time $t$ as

$$\text{X/O} = L^{-3/(2-\beta)} \exp\{ (1-1/L)\, \tau/t \}\ r^S, \tag{1}$$

where $L = \lambda_X / \lambda_O = r^\alpha = ( (A_x/Q_x) / (A_O/Q_O) )^\alpha$ and $\tau = 3R^2/ [\, \lambda_O (2-\beta)^2\, v]$ for particles of speed $v$. Since we compare different ions at the same velocity, their rigidities can be replaced by their corresponding values of $A/Q$. The factor $r^S$ is included to represent any $A/Q$-dependent power-law enhancement at the source. For shock acceleration of impulsive suprathermal ions, it describes the power-law enhancement of the seed particles expected from acceleration in impulsive SEP events (*e.g.* Drake *et al.*, 2009). For shock acceleration of the ambient coronal material, $S = 0$.

To simplify Equation 1, we can achieve a power-law approximation if we expand $\log x = (1-1/x) + (1-1/x)^2/2 + \ldots$ (for $x > \frac{1}{2}$). Using only the first term to replace $1-1/L$ with $\log L$ in Equation (1), we have

$$\text{X/O} \approx L^{\tau/t - 3/(2-\beta)}\, r^S, \tag{2}$$

as an expression for the power-law dependence of enhancements on $A/Q$ for species X.

More generally, we can write Equation 2 in the form $\text{X/O} = r^p$, where the exponent $p$ is linear in the variable $1/t$, so that

$$p = \alpha\, \tau / t + S - 3\alpha / (2-\beta), \tag{3}$$

the average parameters are directly measurable from the time behavior of SEP-abundance observations (Reames, 2016b). Fits to the time behavior of typical gradual SEP events show that late in events where impulsive material is not present $p \approx -1$ to $-2$ (Reames, 2016b), while the average power-law for impulsive events suggests $S \approx 3$ (Reames, Cliver, and Kahler, 2014a, Mason, *et al.* 2004). The ions in small impulsive events generally propagate scatter free (Mason *et al.*, 1989), but in the more intense gradual events any shock-accelerated impulsive suprathermal ions are scattered, so that the power for the heavy-element enhancements is reduced (Reames, 2016b).

   Thus we can expect that the enhancements will vary as a power law in *A/Q,* but the pattern of the average value of *Q* for each element depends upon the source plasma temperature *T* of the ions.  Figure 1 shows two examples of observed abundance enhancements for SEP events matched against regions of the theoretical plots of *A/Q versus T* based upon Mazzotta *et al.* (1998).

**Figure 1**.  Observed element enhancements (*left*) are compared with theoretical plots of *A/Q versus T* (*right*) for 8-hr periods in 14 November 1998 (*upper panels*) and 22 May 2013 (*lower panels*) SEP events. Temperatures are chosen that best match the observed groupings of elements (Reames 2016a).

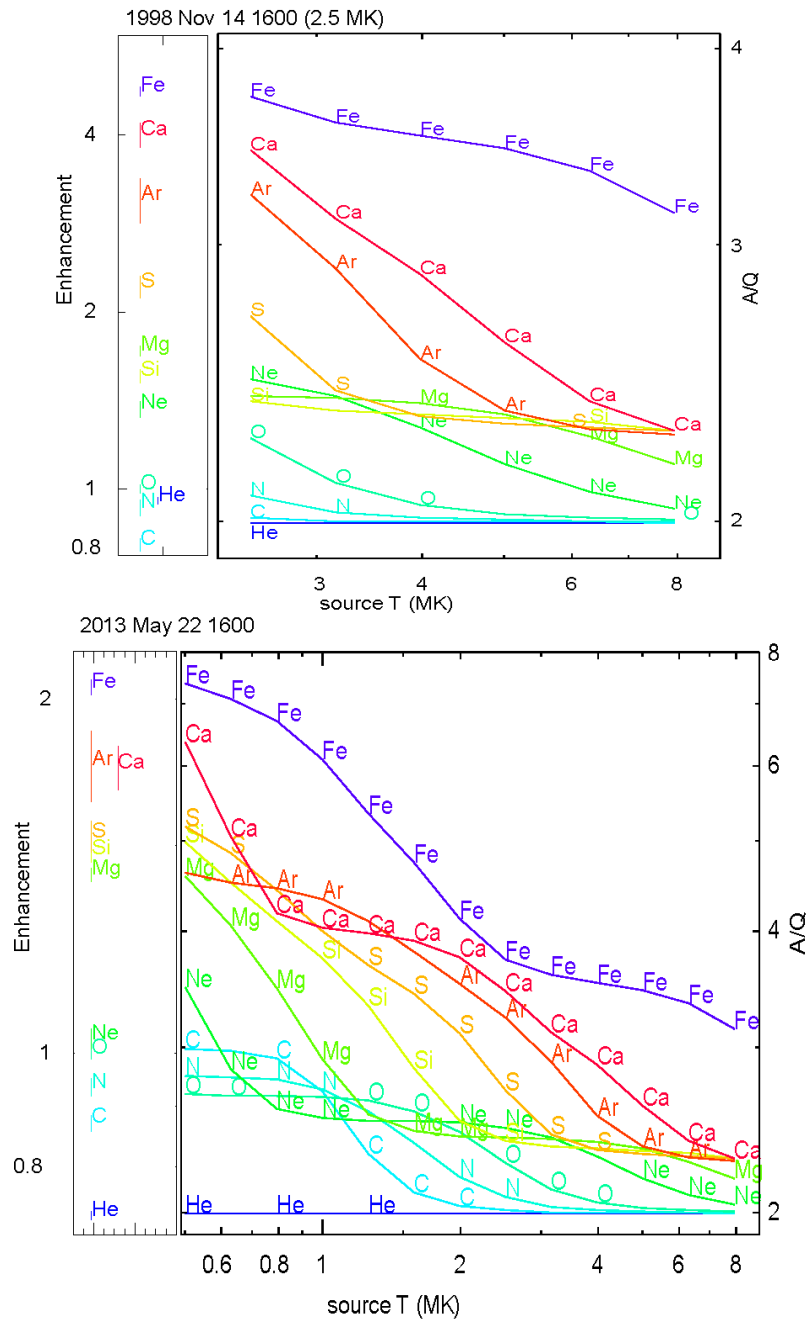

For the event at $T \approx 2.5$ MK in the upper left panel of Figure 1, the almost fully ionized elements C, N, and O occur grouped with He, while Ne, Mg, and Si, with two orbital electrons, are grouped at an interval above them. For the example in the lower panels, with $T \approx 0.7$ MK, partially ionized C, N, and O have moved well above He and are now grouped with Ne, while Mg and Si have moved far above Ne and have joined S approaching Ar and Ca. We will see that the $\chi^2$ values of fits can be quite sensitive to these variations.

Given a temperature and the associated values of $A/Q$, we can determine, by least-squares fitting, the best fit line and the associated value of $\chi^2/m$, where $m$ is the number of degrees of freedom. Repeating this fitting for every $T$ of interest, we can plot $\chi^2/m$ *versus* $T$ (*e.g.* Figure 2d) to choose the most likely temperature and fit line (Reames 2016a, 2016b, 2018b). In the present article, we include only the elements $Z \geq 6$ in the fits, and compare that best fit line with He and extrapolated to H. In the fits an error of 15% is assumed for random abundance variations in addition to the statistical errors.

## 3. Low-*T* Events and Decreasing Powers of *A/Q*

Guided by the power-law fits of gradual events for $Z \geq 2$ by Reames (2016a), we first examine events with $T < 2$ MK, which seem to represent shock acceleration of the ambient solar coronal plasma.

Figure 2 shows the analysis of the event of 24 August 1998 from a source at longitude E09 on the Sun. The extrapolated values of the least-squares power-law fits for elements with $Z \geq 6$ for the six 8-hr time intervals agree rather well with the measured enhancement for H plotted at $A/Q = 1$. In this event, the abundances of the elements with $Z \geq 6$ actually predict the abundances of H and He.

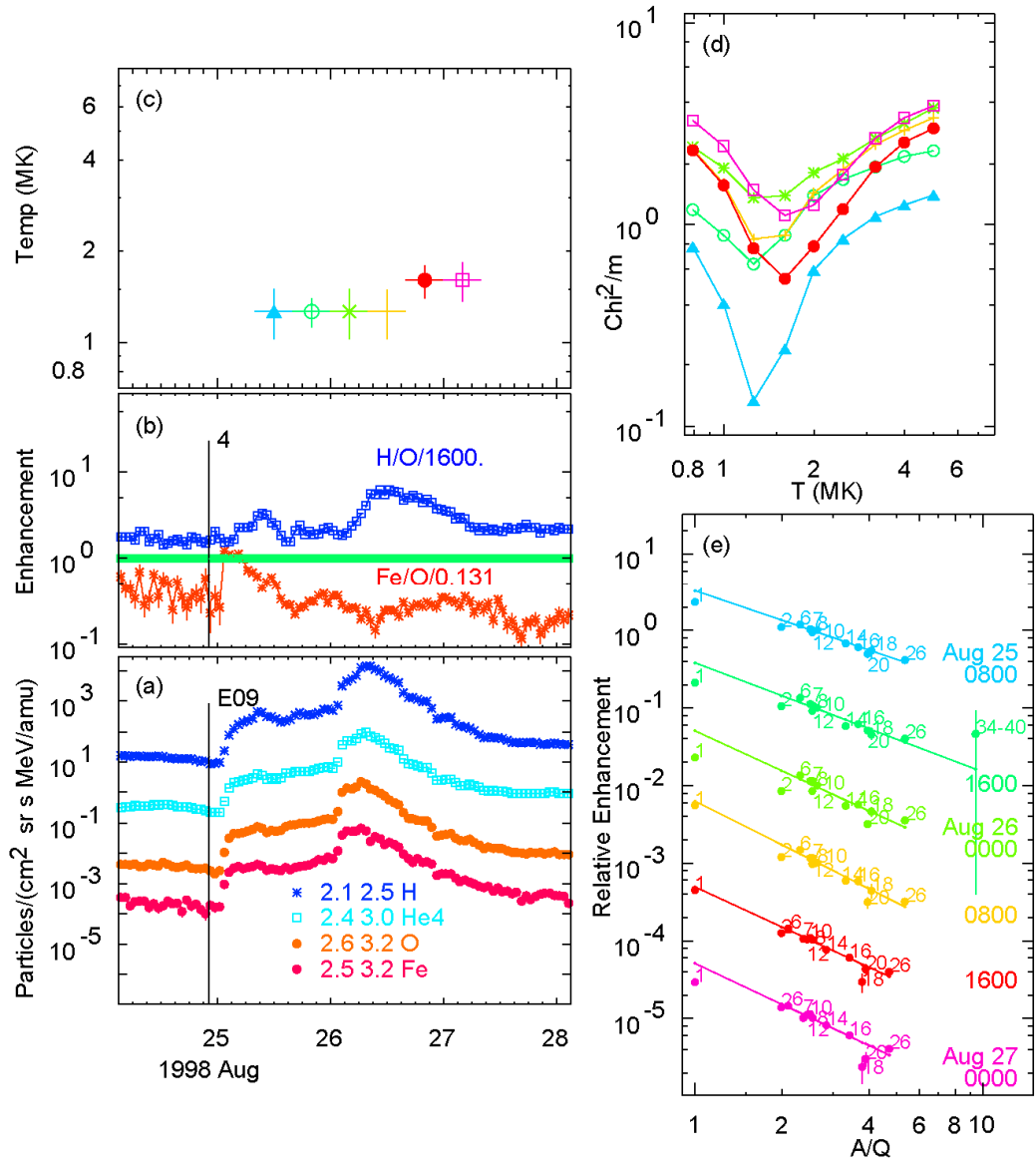

**Figure 2**. (a) Intensities of H, He, O, and Fe *versus* time. (b) Normalized abundance enhancements H/O and Fe/O *versus* time. (c) Best-fit temperatures are shown *versus* time for the 24 August 1998 SEP event. (d) Shows $\chi^2/m$ *versus* T for each 8-hr interval. (e) Shows enhancements, labeled by Z, *versus* A/Q for each 8-hr interval shifted ×0.1, with best-fit power law for elements with $Z \geq 6$ extrapolated down to H at $A/Q = 1$. Colors correspond for the six intervals in (c), (d), and (e) and symbols in (c) and (d); times are also listed in (e). Event onset is flagged with solar longitude in (a) and event number from Reames (2016a) in (b).

Figure 3 shows the event of 23 January 2012 from a source at W21 on the Sun with an associated CME speed of 2175 km s$^{-1}$. The enhancements *versus A/Q* are quite flat during the early periods so T is poorly determined. Here again the abundance of H is fairly well predicted by the abundances of elements with $Z \geq 6$.

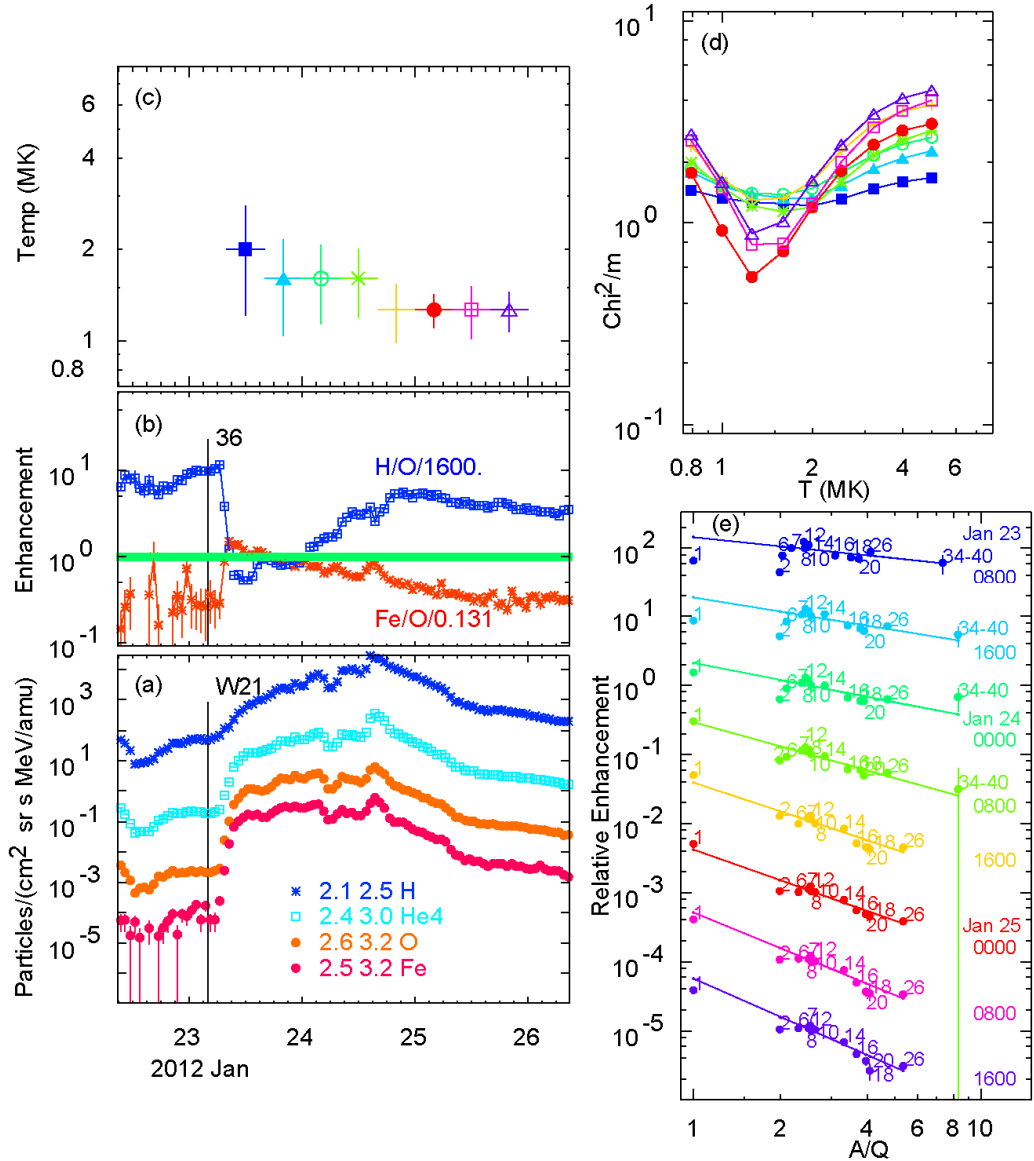

**Figure 3.** (a) Intensities of H, He, O, and Fe *versus* time. (b) Normalized abundance enhancements H/O and Fe/O *versus* time. (c) Best-fit temperatures *versus* time for the 23 January 2012 SEP event. (d) Shows $\chi^2/m$ *versus* $T$ for each 8-hr interval. (e) Shows enhancements, labeled by $Z$, *versus* $A/Q$ for each 8-hr interval shifted ×0.1, with best-fit power law for elements with $Z \geq 6$ extrapolated down to H at $A/Q = 1$. Colors correspond for the eight intervals in (c), (d), and (e) and symbols in (c) and (d); times are also listed in (e). Event onset is flagged with solar longitude in (a) and event number from Reames (2016a) in (b).

Figure 4 shows analysis of the first two of the Halloween events on 26 and 28 October 2003 from W38 and E08 with CME speeds of 1537 and 2459 km s$^{-1}$, respectively. Extrapolation of the least-squares fits for $Z \geq 6$ to $A/Q = 1$ fit H rather well, even when the slope reverses for the last interval in Figure 4e. These events have been studied pre-viously for their protons (Mewaldt *et al.* 2005) and their heavier ions (Cohen *et al.* 2005) separately.

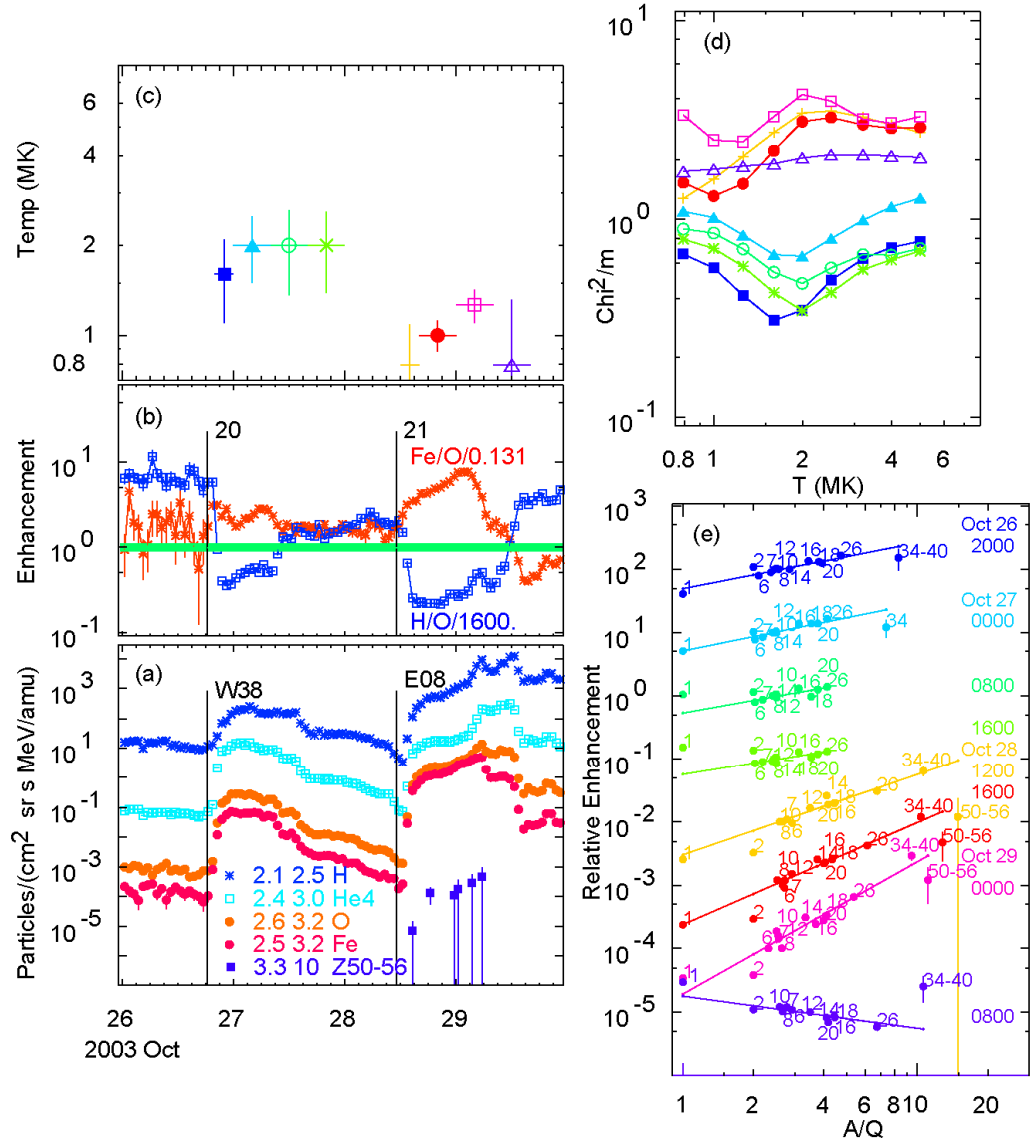

**Figure 4.** (a) Intensities of H, He, O, Fe, and $50 \leq Z \leq 56$ ions *versus* time. (b) Normalized abundance enhancements H/O and Fe/O *versus* time. (c) Derived source temperatures *versus* time, all for the 26 and 28 October 2003 SEP events. (d) Shows $\chi^2/m$ *versus* $T$ for each 8-hr interval. (e) Shows enhancements, labeled by $Z$, *versus* $A/Q$ for each 8-hr interval shifted ×0.1, with best-fit power law for elements with $Z \geq 6$ extrapolated down to H at $A/Q = 1$. Colors correspond for the eight intervals in (c), (d), and (e) and symbols in (c) and (d); times are also listed in (e). Event onsets are flagged with solar longitude in (a) and event number from Reames (2016a) in (b).

# 4. High-*T* Events with Increasing Powers of *A/Q*

H seemed to fit the abundance pattern of the preceding events.  However, there is another class of events where it does not; examples have already been seen for impulsive SEP events (Reames 2019b).   Events with source plasma temperatures in the 2.5–3.2 MK range and positive power-law increase with $A/Q$ have been associated with shock reacceleration of impulsive suprathermal ions (Reames 2016a, 2016b, 2017c, 2018b).   In our sample of 45 gradual SEP events with measurable source plasma temperatures, 11 (24%) were of this class (Reames 2016a).  These events have very small variations in abundance ratios, especially C/He and O/C, compared with impulsive SEP events (Reames 2016b).  Since He, C, and O are nearly fully ionized at these temperatures, the small variations suggest that these "hot" gradual events have averaged abundances over a seed population consisting of many small impulsive SEP contributions (Reames 2016b).

Figure 5 is an example of an event with positive power-law slope and $T \approx 2.5$ MK for an event from W63 on the Sun with a CME speed of 1556 km s$^{-1}$.  These properties suggest the presence of impulsive suprathermal ions with S > 0 in the seed population. The proton enhancement in this event is an order of magnitude higher than expected from the extrapolation of the fit for the ions with $Z \geq 6$.  This behavior was seen recently for the impulsive events where there is additional reacceleration of ions by shock waves (Reames 2019b) and we will find this behavior common for many gradual SEP events with positive powers of $A/Q$ and with $T > 2$ MK, since they are dominated by reaccelerated impulsive SEP ions.

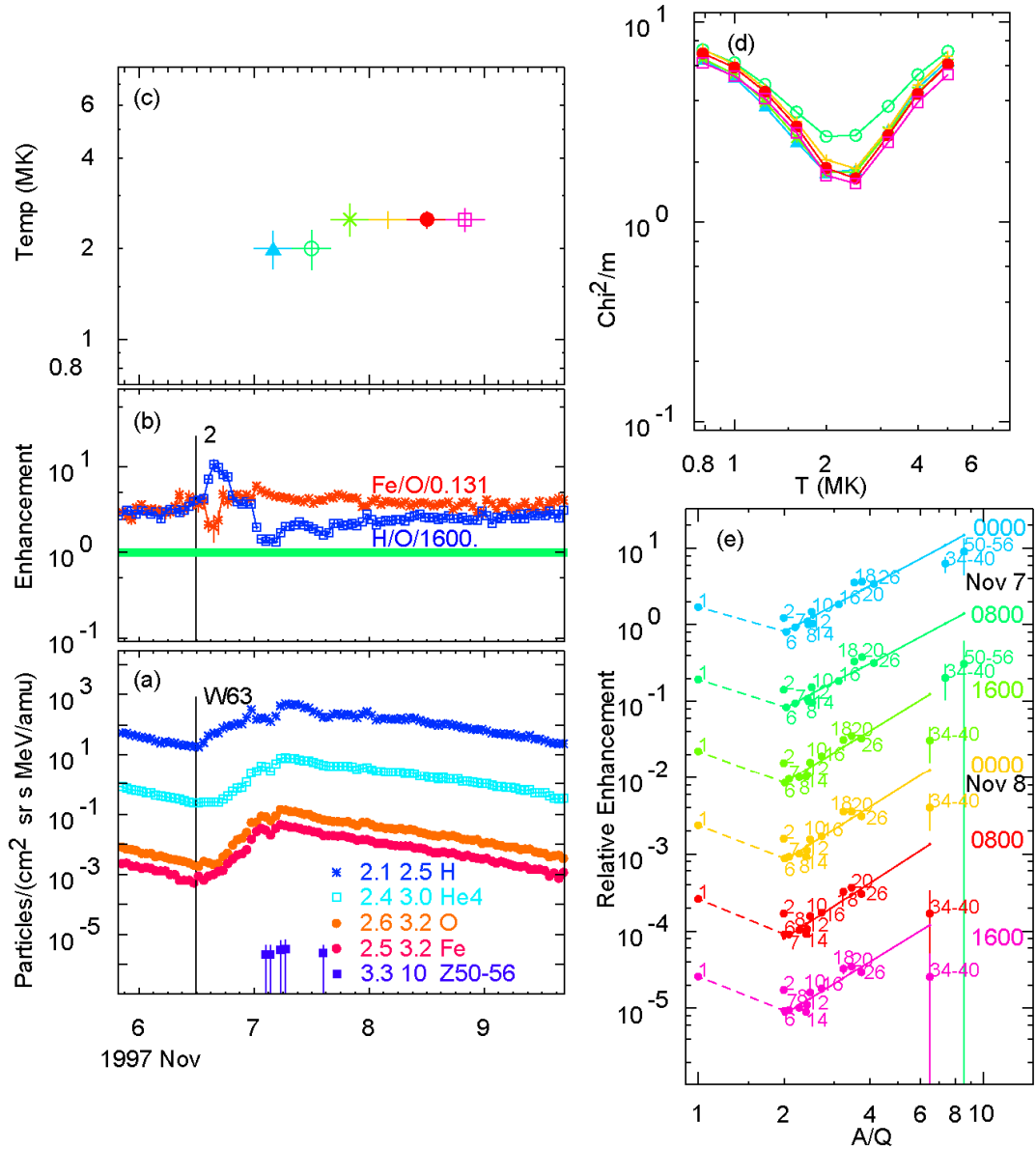

**Figure 5**. (a) Intensities of H, He, O, Fe, and $50 \leq Z \leq 56$ ions *versus* time. (b) Normalized abundance enhancements H/O and Fe/O *versus* time. (c) Temperatures *versus* time for the 6 November 1997 SEP event. (d) Shows $\chi^2/m$ *versus* $T$ for each 8-hr interval. (e) Shows enhancements, labeled by $Z$, *versus* $A/Q$ for each 8-hr interval shifted ×0.1, with best-fit power law for elements with $Z \geq 6$ (*solid*) joined to H by *dashed lines*. Colors correspond for the six intervals in (c), (d), and (e) and symbols in (c) and (d); times are also listed in (e). Dashed lines join H with its associated elements in (e). Event onset is flagged with solar longitude in (a) and event number from Reames (2016a) in (b).

Figure 6 shows another example of an event with a positive power-law slope and $T \approx 3$ MK, that of 14 November 1998 from a source behind the limb at W120.

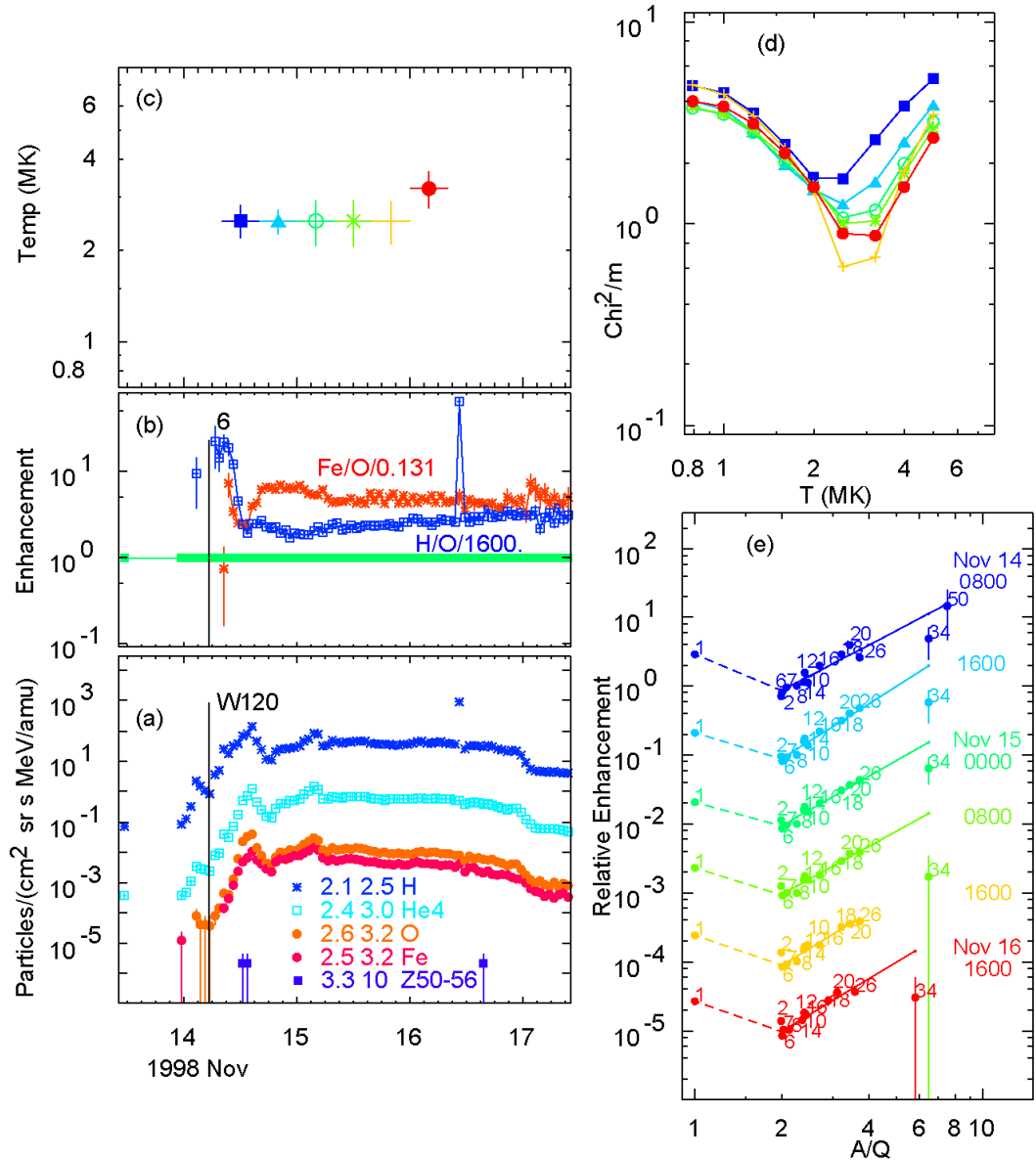

**Figure 6.** (a) Intensities of H, He, O, Fe, and $50 \leq Z \leq 56$ ions *versus* time. (b) Normalized abundance enhancements H/O and Fe/O *versus* time. (c) Temperatures *versus* time for the 14 November 1998 SEP event. (d) Shows $\chi^2/m$ *versus* $T$ for each 8-hr interval. (e) Shows enhancements, labeled by $Z$, *versus* $A/Q$ for each 8-hr interval shifted ×0.1, with best-fit power law for elements with $Z \geq 6$ (*solid*) joined to H by *dashed lines*. Colors correspond for the six intervals in (c), (d), and (e) and symbols in (c) and (d); times are also listed in (e). Dashed lines join H with its associated elements in (e). Event onset is flagged with solar longitude in (a) and event number from Reames (2016a) in (b).

For comparison, we show in Figure 7 the next event in the series of Halloween events, immediately following the events in Figure 4. The source of this event is a CME with a speed of 2029 km s⁻¹ from longitude W02 on the Sun on 29 October 2003.

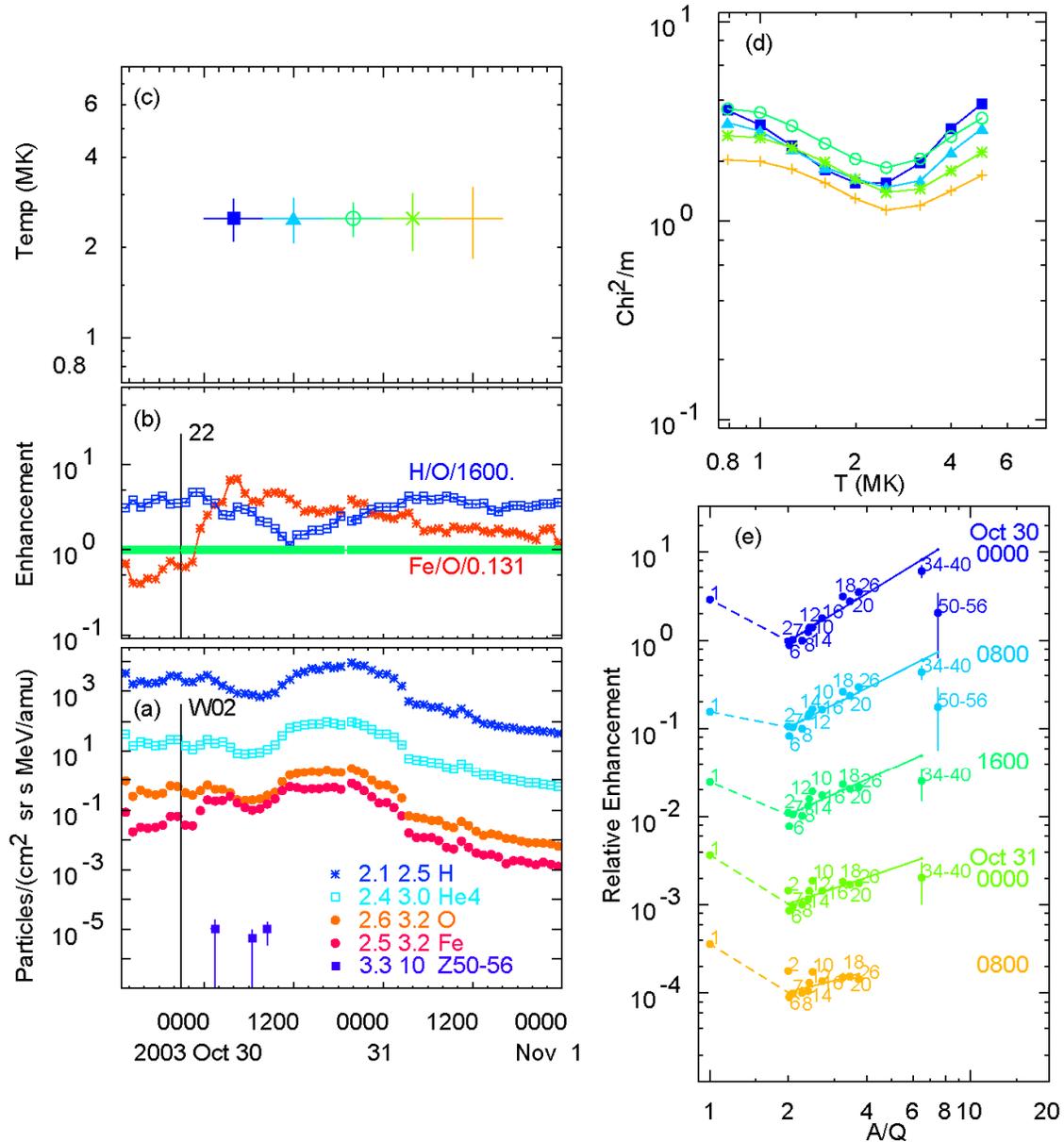

**Figure 7.** (a) Intensities of H, He, O, Fe, and $50 \leq Z \leq 56$ ions *versus* time. (b) Normalized abundance enhancements H/O and Fe/O *versus* time. (c) Temperatures are shown *versus* time for the 29 October 2003 SEP event. (d) Shows $\chi^2/m$ *versus* T for each 8-hr interval. (e) Shows enhancements, labeled by $Z$, *versus* A/Q for each 8-hr interval shifted ×0.1, with best-fit power law for elements with $Z \geq 6$ (*solid*) joined to H by *dashed lines*. Colors correspond for the five intervals in (c), (d), and (e) and symbols in (c) and (d); times are also listed in (e). Dashed lines join H with its associated elements in (e). Event onset is flagged with solar longitude in (a) and event number from Reames (2016a) in (b).

Unlike its immediate predecessors (shown in Figure 4) the event in Figure 7 shows a 2.5 MK seed population of impulsive SEP ions with an order-of-magnitude ex-cess of protons.  This event is not influenced by the properties of its immediate predecessor in Figure 4.

# 5. Other Variations

From the foregoing one might conclude that hydrogen enhancements agree with those of other elements when the shock waves accelerate ambient coronal material and H exceeds expectations by a factor of about 10 when impulsive SEP ions are reaccelerated.  That does describe most gradual SEP events.  However, there are some events with combinations of behavior that are more complex.

Figure 8 shows analysis of the event of 2 April 2001 driven by a CME with a speed of 2505 km s$^{-1}$ from W78 on the Sun.  Abundances early in the event indicate a temperature of 2.5 MK, but the proton excess is quite modest, perhaps a factor of two above expectations.  By the middle of the event the temperature rises slightly to 3.2 MK and the factor-of-ten proton excess is seen.

The final three periods shown in Figure 8e illustrate that temperature determination becomes impossible when the abundance enhancements are flat, independent of $A/Q$. We can measure temperatures when there is either increasing or decreasing dependence on $A/Q$, but we cannot determine $A/Q$ or $T$ when there is no variation and the enhancements are independent of $A/Q$.  Nevertheless, it is clear that the proton enhancements are a factor of ten above the other elements here as well.

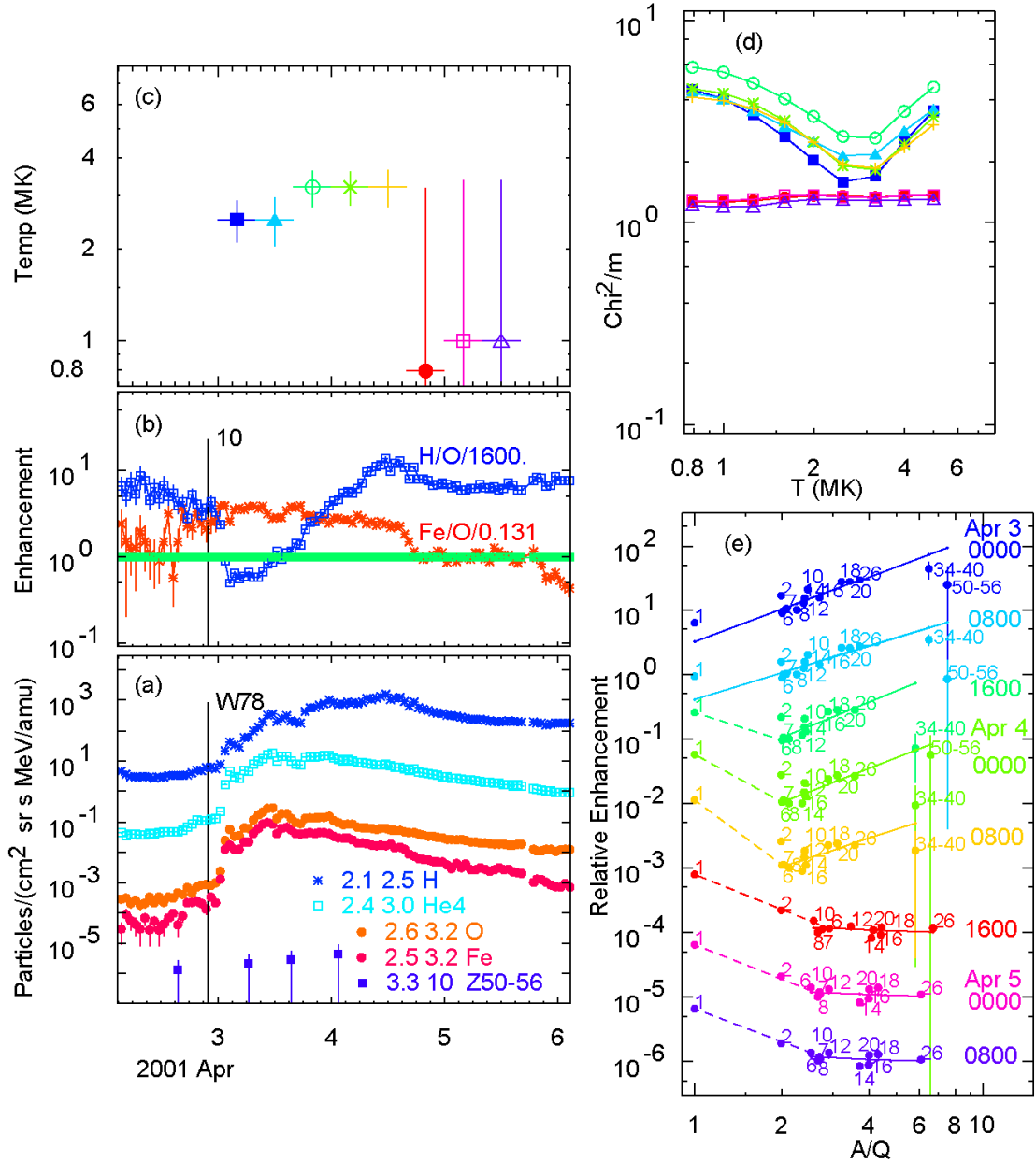

**Figure 8**. (a) Intensities of H, He, O, Fe, and $50 \leq Z \leq 56$ ions *versus* time. (b) Normalized abundance enhancements H/O and Fe/O *versus* time. (c) Temperatures are shown *versus* time for the 2 April 2001 SEP event. (d) Shows $\chi^2/m$ *versus* $T$ for each 8-hr interval. (e) Shows enhancements, labeled by $Z$, *versus* $A/Q$ for each 8-hr interval shifted ×0.1, with best-fit power law for elements with $Z \geq 6$ extrapolated down to H at $A/Q = 1$ (*solid*) or joined to H by *dashed lines*. Colors correspond for the eight intervals in (c), (d), and (e) and symbols in (c) and (d); times are also listed in (e). Dashed lines join H with its associated elements in (e). Event onset is flagged with solar longitude in (a) and event number from Reames (2016a) in (b).

Figure 9 shows the event of 29 September 2013 from W33 with a CME speed of 1179 km s$^{-1}$. This event shows the classic evolution where Fe/O decreases with time.

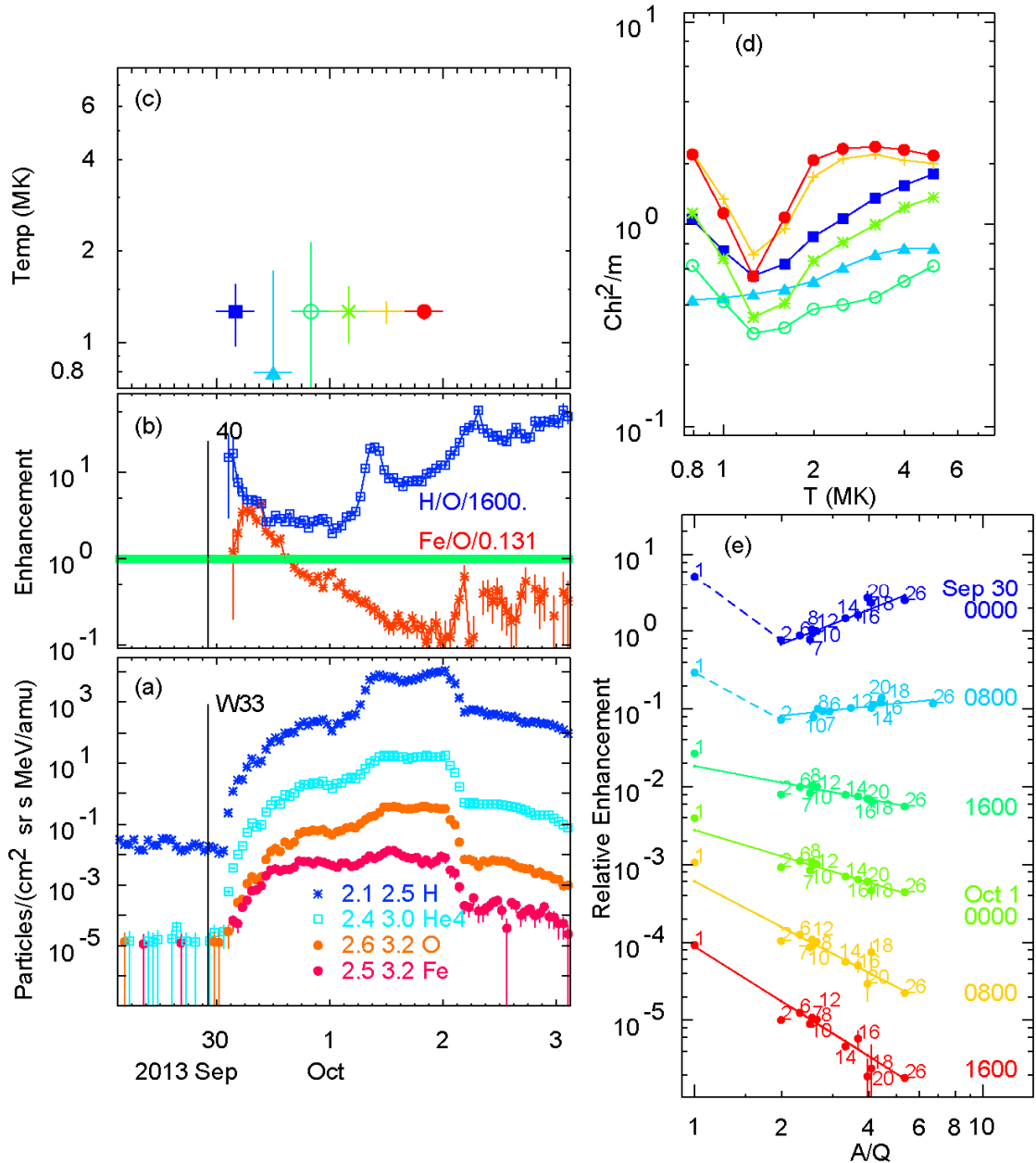

**Figure 9**. (a) Intensities of H, He, O, and Fe ions *versus* time. (b) Normalized abundance enhancements H/O and Fe/O *versus* time. (c) Temperatures are shown *versus* time for the 29 September 2013 SEP event. (d) Shows $\chi^2/m$ *versus* $T$ for each 8-hr interval. (e) Shows enhancements, labeled by $Z$, *versus* $A/Q$ for each 8-hr interval shifted ×0.1, with best-fit power law for elements with $Z \geq 6$ extrapolated down to H at $A/Q = 1$ (*solid*) or joined to H by *dashed lines*. Colors correspond for the six intervals in (c), (d), and (e) and symbols in (c) and (d); times are also listed in (e). Dashed lines join H with its associated elements in (e). Event onset is flagged with solar longitude in (a) and event number from Reames (2016a) in (b).

The event in Figure 9 shows a consistent temperature of 1.5 MK indicating a lack of hotter material with impulsive suprathermal ions, yet the proton enhancement is sig-nificantly above expectation early, when the enhancements are rising with $A/Q$, but are well within expectations when the enhancements are falling with $A/Q$.

Does the large proton excess depend upon the presence of impulsive material or on the slope of the power law? Figure 9e suggests that only the slope matters, but Figure 8e shows predicted proton abundances for both positive slope and $T = 2.5$ MK.

# 6. Proton Excess and Distributions in $A/Q$ and $T$

We have seen several examples of events that show a typical behavior of proton enhancement in gradual SEP events and a few events that show an atypical behavior. What remains is to examine the probability distributions for a wide range of cases. Since the behavior can vary with time during an event we must look at the distributions of the basic 8-hr averages, nearly 400 intervals within the SEP events originally studied by Reames (2016a).

The upper panel of Figure 10 shows the distribution of proton enhancements during the 8-hr intervals as a function of the slope, or power of $A/Q$, of the fit. Clearly, the peak of the distribution lies at a negative slope near -1, where most intervals have a proton enhancement within a factor of about three of having the value correctly predicted by the ions with $Z \geq 6$ or modest excesses. Grouping the intervals by temperature, in the lower panels of Figure 10, shows that intervals with $T \geq 2.5$, on the right, identify with the positive slope involving reaccelerated impulsive ions with typical power-of-ten proton excesses. Lower temperatures of ambient coronal plasma in the lower left panel have a much better chance of predicting the proton intensity using only the $Z \geq 6$ ions, but there are some event intervals with positive slope that usually occur early in these events. In the original temperature distribution of events (Reames 2016a), which is unchanged here, 24 % (11/45) of the events were in the 2.5–3.2 MK interval and 69 % (31/45) had $T \leq 1.6$ MK. The proton excesses tend to follow that division, not because of the temperature itself but, more likely, because of the associated distribution of positive and negative slope of $A/Q$.

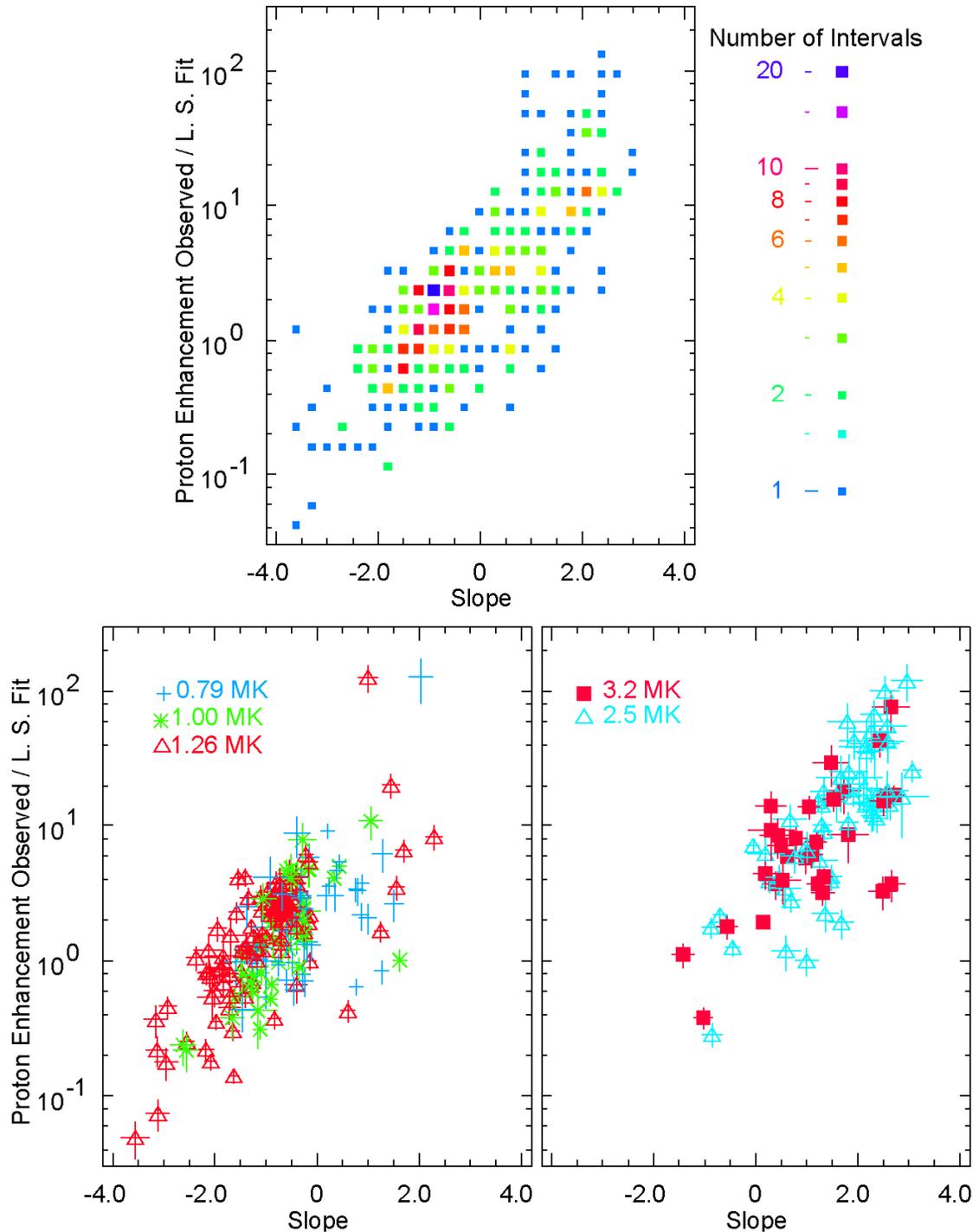

**Figure 10**. In all panels, the enhancement of H relative to that expected from the least-squares (L.S.) power-law fit of elements $Z \geq 6$ is shown *versus* the "slope" or power of $A/Q$ from the fit of elements $Z \geq 6$. The *upper panel* shows a histogram of the distribution of all 398 8-hr intervals in this space with symbol color and size showing the number at each location. The *lower left panel* shows the distribution of intervals with $T =$ 0.79, 1.0, and 1.26 MK. The *lower right panel* shows the distribution of intervals with $T =$ 2.5 and 3.2 MK.

# 7. Discussion

The abundances we study depend upon two underlying factors. First, elements suffer the FIP effect as they are brought into the corona to produce the reference SEP abundances tabulated in Appendix A. Second, their acceleration and transport depend, not only upon particle rigidity, but approximately as a power law on rigidity. Similar processes apply, not only to SEPs, but even to energetic ions in corotating interaction regions where shock waves accelerate ions, from the solar wind in this case, to produce enhancements that usually decline with $A/Q$ (Reames, 2018c) like most SEP events here. All ions experience this process, but the protons in shock acceleration have a dual role; they are not only accelerated like other ions, but they must generate waves as they stream away so as to scatter subsequent ions back and forth across the shock (Bell 1978; Lee, 1983, 2005; Ng and Reames, 2008; Desai and Giacalone, 2016). Are extra protons sometimes required for the latter task? If we adjust the abundance normalization from O to H, events with formerly "excess protons" are now seen as unable to accelerate the expected number of heavy ions given the available protons and the wave spectrum they have produced. However, we might not be surprised to find excess protons needed early in events when the shocks are strong and resonant waves are just forming, but less so later as the shocks weaken and the pattern of waves is established (see also Reames, Ng, and Tylka, 2000).

When we compare ions at the same velocity, as we do for studies of abundance enhancements, the ions interact with different parts of the proton-generated wave spectrum. For example, for 2.5 MeV protons the wave spectrum is approximately generated by streaming 2.5 MeV protons, for 2.5 MeV amu$^{-1}$ He, C, or O with $A/Q = 2$ the spectrum is generated by streaming 10 MeV protons, for 2.5 MeV amu$^{-1}$ Fe at $A/Q = 4$ the spectrum is generated by streaming 39 MeV protons, and for 2.5 MeV amu$^{-1}$ heavy ions with $A/Q = 10$ the spectrum is generated by streaming 224 MeV protons. Thus the proton spectrum and the slope of the $A/Q$-dependence of heavy ion enhancements are related. Can a proton spectrum, which usually declines with energy, produce enough waves to accelerate heavier ion abundances that increase with resonant rigidity? For positive slopes in $A/Q$, perhaps we should compare 39 MeV protons with 2.5 MeV amu$^{-1}$ Fe. Unfortunately, 39 MeV proton data are not available during these events.

Proton spectral shape can be important, and any downward breaks in the proton spectrum will especially tend to depress heavy ions perhaps such as may be occurring for the $50 \leq Z \leq 56$ ions in Figures 7, and 8. Breaks can also disrupt the power-law lower in $A/Q$ as seen by the break at Mg in the multi-spacecraft study of the SEP event of 23 January 2012 (Reames 2017b).

Do the excess protons come from the same ion population as the heavy ions? Recently, Reames (2019b) suggested that two components contributed to the shock acceleration in impulsive SEP events with significant excess protons, the heavy ions may come from residual impulsive suprathermal ions with positive power of $A/Q$, while the protons come from the ambient plasma with a negative power of $A/Q$ (see Figure 9 in Reames 2019b). However, excess protons are seen rather often and in similar amounts, making this dual-source explanation less probable, since two components might be expected to produce a larger range of relative variation. Also, two components would not explain the event in Figure 9 where a proton excess and positive power of $A/Q$ coexists for ions with an ambient 1.5 MK source plasma temperature. Apparently, more protons are required to support shock conditions that can produce or can maintain a strong positive power of $A/Q$ in the enhancements of heavy elements, even when injected abundances already have a positive power of $A/Q$ initially. However, cases do exist where positive powers of $A/Q$ extrapolate to predict the observed proton enhancement (Figures 4 and 8).

We summarize the general behavior of gradual SEP events as follows:

i) Most time intervals in most gradual SEP events (> 60 %) have source temperatures $0.7 \leq T \leq 1.6$ MK and abundances that decrease as a power of $A/Q$. Proton enhancements are predicted by the power-law pattern of the other elements.

ii) Time periods during those events (> 20%) dominated by pre-accelerated impulsive ions, with $T > 2$ MK and abundances that increase as a power of $A/Q$, usually have proton enhancements a factor of about ten above the expected values.

When we consider impulsive and gradual SEP events together, three situations can exist. Proton abundances can be consistent with the expectations of a power law in $A/Q$ in (i) "pure" small Fe-rich impulsive SEP events (Reames, 2019b) and in (ii) "pure" large gradual SEP events with decreasing powers of *A/Q* with few impulsive ions. However, proton abundances significantly exceed expectations primarily in (iii) "compound" SEP events where shock waves reaccelerate impulsive SEP material. This new three-way behavior of proton abundances can help identify the physical processes involved. Thus, Fe-rich SEPs with expected proton abundances were probably accelerated in a magnetic reconnection region without involvement of shocks, while Fe-rich SEPs with a 10-fold excess of protons were probably reaccelerated by a shock wave. This was a powerful new insight for impulsive SEP events.

Is there a difference between the impulsive events where proton intensities are boosted by the narrow shock wave from their associated CME and the gradual events where fast, wide shock waves predominantly sample impulsive suprathermal seed populations? Perhaps there is, since the shock in the former samples a single impulsive injection while the latter can average over the cumulative residual ions from many impulsive events, greatly averaging and reducing the event-to-event abundance variations as discussed by Reames (2016b).

In contrast to H , He shows substantial source abundance variations $30 \leq He/O \leq 100$ in both impulsive and gradual events (Reames, 2017c, 2019a) with a few impulsive SEP events showing extreme suppression of $He/O \approx 2$ (Reames, 2019a). These variations in He may result from FIP-related inefficient ionization of He during transport up into the corona (Laming, 2009). H and He do not share the same behavior; this is why we study H/O and He/O rather than He/H. In view of the comparisons of other elements (*e.g.* Reames, 2018a), it is not surprising that variations of H and He in SEP events also seem to differ completely from variations of H and He in the solar wind.

Further progress in understanding proton abundances may depend upon theory and modeling that includes other parameters that can affect acceleration and transport. Do the current models of SEP acceleration and transport support the observed *A/Q* dependence? Do they predict excess protons? The goal of this article has been to examine SEP events from a new perspective and to show the range of variations and suggest the probable cause. More detailed theoretical studies relating proton spectra to *A/*Q dependence are needed.

## Disclosure of Potential Conflicts of Interest

The author declares he has no conflicts of interest.

# Appendix A: Reference Abundances of Elements

The average element abundances in gradual SEP events are a measure of the coronal abundances sampled by SEP events (reference gradual SEPs in Table 1).  They differ from photospheric abundances (Table 1) by a factor which depends upon FIP (*e.g.* Reames 2018a; 2018b).  Ion "enhancements" are defined as the observed abundance of a species, relative to O, divided by the reference abundance of that species, relative to O.

**Table 1** Reference gradual SEP abundances, photospheric, and impulsive SEP abundances are shown for various elements.

| | Z | FIP [eV] | Photosphere[1] | Reference Gradual SEPs[2] | Avg. Impulsive SEPs[3] |
|---|---|---|---|---|---|
| H | 1 | 13.6 | $(1.74\pm0.17)\times10^{6}$ * | $(1.6\pm0.2)\times10^{6}$ | - |
| He | 2 | 24.6 | 156000±7000 | 57000±5000 | 53000±3000 ** |
| C | 6 | 11.3 | 550±76 * | 420±10 | 386±8 |
| N | 7 | 14.5 | 126±35 * | 128±8 | 139±4 |
| O | 8 | 13.6 | 1000±161 * | 1000±10 | 1000±10 |
| Ne | 10 | 21.6 | 210±54 | 157±10 | 478±24 |
| Mg | 12 | 7.6 | 64.5±9.5 | 178±4 | 404±30 |
| Si | 14 | 8.2 | 61.6±9.1 | 151±4 | 325±12 |
| S | 16 | 10.4 | 25.1±2.9 * | 25±2 | 84±4 |
| Ar | 18 | 15.8 | 5.9±1.5 | 4.3±0.4 | 34±2 |
| Ca | 20 | 6.1 | 4.0±0.7 | 11±1 | 85±4 |
| Fe | 26 | 7.9 | 57.6±8.0 * | 131±6 | 1170±48 |
| Zn-Zr | 30-40 | - | - | 0.04±0.01 | 2.0±0.2 |
| Sn-Ba | 50-56 | - | - | 0.0066±0.001 | 2.0±2 |
| Os-Pb | 76-82 | - | - | 0.0007±0.0003 | 0.64±0.12 |

[1]Lodders, Palme, and Gail (2009), see also Asplund *et al.* (2009).

* Caffau *et al.* (2011).

[2] Reames (1995, 2014, 2017a).

[3] Reames, Cliver, and Kahler (2014a)

** Reames (2019a).